\documentclass[10pt]{article}
\usepackage{amsmath, amssymb, xcolor, slashed, latexsym}
\usepackage[a4paper, left=2.5cm, right=2.5cm, top=3.5cm, bottom=3cm]{geometry}
\usepackage[hidelinks]{hyperref}
\usepackage{tensor}
\usepackage{tikz-feynman}
\usepackage{hyperref}
\usepackage{dsfont}
\usepackage{simplewick}
\usepackage{color}
\hypersetup{
	colorlinks=true,
	linkcolor=blue,
	citecolor=magenta,
}

\makeatletter
\DeclareFontFamily{OMX}{MnSymbolE}{}
\DeclareSymbolFont{MnLargeSymbols}{OMX}{MnSymbolE}{m}{n}
\SetSymbolFont{MnLargeSymbols}{bold}{OMX}{MnSymbolE}{b}{n}
\DeclareFontShape{OMX}{MnSymbolE}{m}{n}{
	<-6>  MnSymbolE5
	<6-7>  MnSymbolE6
	<7-8>  MnSymbolE7
	<8-9>  MnSymbolE8
	<9-10> MnSymbolE9
	<10-12> MnSymbolE10
	<12->   MnSymbolE12
}{}
\DeclareFontShape{OMX}{MnSymbolE}{b}{n}{
	<-6>  MnSymbolE-Bold5
	<6-7>  MnSymbolE-Bold6
	<7-8>  MnSymbolE-Bold7
	<8-9>  MnSymbolE-Bold8
	<9-10> MnSymbolE-Bold9
	<10-12> MnSymbolE-Bold10
	<12->   MnSymbolE-Bold12
}{}

\let\llangle\@undefined
\let\rrangle\@undefined
\DeclareMathDelimiter{\llangle}{\mathopen}%
{MnLargeSymbols}{'164}{MnLargeSymbols}{'164}
\DeclareMathDelimiter{\rrangle}{\mathclose}%
{MnLargeSymbols}{'171}{MnLargeSymbols}{'171}
\makeatother

\newcommand{\N}{\mathds N}

\newcommand{\A}{\widetilde A}
\newcommand{\R}{\widetilde R}
\newcommand{\G}{\mathcal{G}}
\newcommand{\rR}{\mathrm{R}}
\def\im{\mathrm{i}}
\def\ep{\mathrm{e}}
\def\pa{\partial}
\def\diff{\mathrm{d}}

\def\sfrac#1#2{{\textstyle\frac{#1}{#2}}}
\def\>{\rangle}
\def\<{\langle}
\def\+{\dagger}

\def\={\ =\ }

\def\und{\quad\textrm{and}\quad}
\def\with{\quad\textrm{with}\quad}

\newcommand{\Tr}{\ensuremath{\operatorname{tr}}}
\newcommand{\drm}{\ensuremath{\mathrm{d}}}
\newcommand{\blambda}{\ensuremath{\bar{\lambda}}}

\newcommand{\intdx}{\ensuremath{\int \drm^4 x\;}}
\newcommand{\bC}{\ensuremath{\bar{C}}}

\newcommand{\partialA}{\ensuremath{\partial\cdot A}}
\newcommand{\FP}{\ensuremath{\Delta_{\mathrm{FP}}[A]}}
\newcommand{\MSS}{\ensuremath{\Delta_{\mathrm{MSS}}[A]}}
\newcommand{\Boxi}{\ensuremath{\Box^{-1}}}
\newcommand{\AT}{\ensuremath{A^{\mathrm{T}}}}
\newcommand{\AL}{\ensuremath{A^{\mathrm{L}}}}

\newcommand{\tA}{\ensuremath{\widetilde{A}}}

\newcommand{\tD}{\ensuremath{\widetilde{D}}}
\newcommand{\tP}{\ensuremath{\widetilde{P}}}
\newcommand{\trD}{\ensuremath{\widetilde{\rD}}}
\newcommand{\tC}{\ensuremath{\widetilde{C}}}
\newcommand{\tbC}{\ensuremath{\widetilde{\bar{C}}}}
\newcommand{\tG}{\ensuremath{\widetilde{G}}}
\newcommand{\tS}{\ensuremath{\widetilde{S}}}
\newcommand{\tF}{\ensuremath{\widetilde{F}}}
\newcommand{\tlambda}{\widetilde{\lambda}}
\newcommand{\tblambda}{\widetilde{\bar{\lambda}}}
\newcommand{\tR}{\widetilde{R}}

\newcommand{\tRl}{\ensuremath{\stackrel{\leftarrow}{\tR}}}

\newcommand{\slA}{\slashed{A}}
\newcommand{\sltA}{\widetilde{\slashed{A}}}
\newcommand{\slAL}{\slashed{\AL}}
\newcommand{\rRl}{\stackrel{\leftarrow}{\rR}}
\newcommand{\nc}{n_{\mathrm{c}}}
\newcommand{\rD}{\mathrm{D}}
\newcommand{\Aone}{A^{(1)}}
\renewcommand{\t}{\times}

\newcommand{\I}{\mathrm{i}}
\newcommand{\SU}{\mathrm{SU}}
\begin{document}
	\title{\bf\huge Construction method for the Nicolai map\\[3pt] in
		 supersymmetric Yang--Mills theories}
	\date{~}
	
	\author{\phantom{.}\\[12pt]
		{\scshape\Large Olaf Lechtenfeld \ and \ Maximilian Rupprecht}
		\\[24pt]
		Institut f\"ur Theoretische Physik\\ and\\ 
		Riemann Center for Geometry and Physics\\[8pt]
		Leibniz Universit\"at Hannover \\ 
		Appelstra{\ss}e 2, 30167 Hannover, Germany
		\\[24pt]
	} 
	
	\clearpage
	\maketitle
	\thispagestyle{empty}
	
	\begin{abstract}
		\noindent\large
		Recently, a universal formula for the Nicolai map in terms of a coupling flow functional differential operator was found. We present the full perturbative expansion of this operator in Yang--Mills theories where supersymmetry is realized off-shell. Given this expansion, we develop a straightforward method to compute the explicit Nicolai map to any order in the gauge coupling. Our work extends the previously known construction method from the Landau gauge to arbitrary gauges and from the gauge hypersurface to the full gauge-field configuration space. As an example, we present the map in the axial gauge to the second order.
	\end{abstract}
	\newpage
	\setcounter{page}{1} 
	
	\section{Introduction}
	\textbf{Nicolai map.} We consider unbroken $\mathcal{N}\=1$ supersymmetric gauge theories in the Wess--Zumino gauge in $D$--dimensional Minkowski spacetime $\mathbb{R}^{1,D-1}\ni x$. The fields $(A,\lambda, D)$ are in the adjoint representation of the gauge group which for simplicity we take to be $\SU(\nc)$ with real antisymmetric structure constants $f^{abc}$ such that
	\begin{equation}\label{eq:structure_constants}
		f^{abc}f^{abd}\=\nc \delta^{cd}\ ,\qquad a,b,\ldots\=1, 2, \ldots , \nc^2{-}1\ .
	\end{equation}
	The Nicolai map \cite{Nic1, Nic2, Nic3} can be defined \cite{LR} as a nonlinear and nonlocal field transformation
	\begin{equation}
		T_g :\ A^a_\mu(x) \ \mapsto\ A'^a_\mu(x;g,A)
	\end{equation}
	of the Yang--Mills fields $A^a_\mu$ ($\mu=0,1,\ldots,D{-}1$),
	invertible at least as a formal power series in~$g$, that satisfies
	\begin{equation}\label{eq:free_f_corr}
		\bigl\llangle X[A] \bigr\rrangle_g \= \bigl\< X[T_g^{-1}A] \bigr\>_0
		\quad\forall\,X \ .
	\end{equation}
	This enables one to compute quantum correlators in the interacting theory by using a free, purely bosonic functional measure. The correlators\footnote{Note that by the vanishing of the vacuum energy in supersymmetric theories, these correlators are normalized such that $\llangle 1 \rrangle_g=\bigl\< 1\bigr\>_g=1$.} are given by
	\begin{equation}
		\begin{aligned}
		\bigl\llangle X[A] \bigr\rrangle_g\ &:=\ \int \rD A \rD \lambda \rD D \rD C \rD \bC\ \ep^{\im S_{\textsc{SUSY}}[A, \lambda, D, C, \bC]} X[A]\ ,\\
		\bigl\< X[A] \bigr\>_g\ &:=\ \int \rD A\ \ep^{\im S_g[A]}\ \MSS\ \FP\  X[A]\ ,
		\end{aligned}
	\end{equation}
	where the latter is obtained from the former by integrating out the auxiliaries $D$ and the anticommuting gauginos $\lambda$ and ghosts $C$, $\bC$. This produces the Faddeev--Popov determinant $\FP$ and the Matthews--Salam--Seiler determinant $\MSS$ (a Pfaffian for Majorana fermions). From the full action
	\begin{equation}
	S_{\textrm{\tiny SUSY}}[A,\lambda,D,C,\bar{C}] \= 
	\int\!\drm^D x\ \Bigl\{ -\sfrac{1}{4} F^a_{\mu\nu}F^{a\;\mu\nu} - \sfrac{1}{2\xi}\G(A)^2
	\ +\ \textrm{fermions}\ +\ \textrm{ghosts}\ +\ \textrm{auxiliaries}\ \Bigr\}\ ,
	\end{equation}
	only the gauge-field action 
	\begin{equation}
		S_g[A] \= \int \drm^D x\ \bigl\{-\sfrac14 F^a_{\mu\nu}F^{a\,\mu\nu}- \sfrac{1}{2\xi}\G(A)^2\bigr\}
	\end{equation}
	remains, with the Yang--Mills field strength
	\begin{equation}
		F^a_{\mu\nu}\=\partial_{\mu} A^a_{\nu}-\partial_{\nu} A^a_{\mu}+gf^{abc}A^b_\mu A^c_\nu\qquad \iff\qquad F_{\mu\nu}\=\partial_\mu A_\nu-\partial_\nu A_\mu +g A_\mu\t A_\nu\ .
	\end{equation}
	The gauge dependence of the Nicolai map enters through the choice of the 't Hooft parameter $\xi$ and the gauge fixing function $\G^a(A)$, for example
	\begin{equation}
	\G^a(A) \= \pa^\mu\! A^a_\mu\ \textrm{(Lorenz gauge)}\quad\textrm{or} \quad \G^a(A) \= n^\mu\! A^a_\mu\ \textrm{(axial gauge)}\ .
	\end{equation}
	For $g\rightarrow 0$, one obtains the free-field correlator in \eqref{eq:free_f_corr},
	\begin{equation}
		\bigl\< X[A] \bigr\>_0\=\sfrac{1}{Z}\int\rD A\ \ep^{\im S_0[A]}X[A]\quad\with \sfrac{1}{Z}\=\Delta_{\mathrm{MSS}}[0]\ \Delta_{\mathrm{FP}}[0]\ .
	\end{equation}

	\noindent\textbf{Universal form.}
	In a recent work \cite{LR}, we demonstrated that the Nicolai map can be written as an ordered exponential
	\begin{equation}
	T_g A \= \overrightarrow{\cal P} \exp \Bigl\{-\!\int_0^g\!\diff h\ R_h[A]\Bigr\}\ A
	\end{equation}
	in terms of a coupling flow differential operator $R_g[A]$, which is defined via
	\begin{equation} \label{localflow}
	\pa_g \bigl\< X[A] \bigr\>_g \= \bigl\< \bigl( \pa_g + R_g[A] \bigr) X[A] \bigr\>_g\ .
	\end{equation}
	If the original local theory possesses an off-shell supersymmetric formulation, then the coupling flow operator can be canonically constructed in any gauge. The procedure is outlined in Section~\ref{sec:coupling_flow}. Our main result is the compact form \eqref{eq:R_compact} and explicit perturbative expansion \eqref{eq:pert_expansion} for
	\begin{equation}\label{eq:series_R}
	R_g[A] \= \sum_{k=1}^\infty g^{k-1} \rR_k[A] \= \rR_1[A] + g\,\rR_2[A] + g^2 \rR_3[A] + \ldots\ ,
	\end{equation}
	which allows us to compute the Nicolai map\footnote{and its inverse, by an equally simple formula} via
	\begin{equation}\label{eq:nmap_coefficients}
	T_g\,A \= \sum_{\bf n} g^n\,c_{\bf n}\,\rR_{n_s}[A]\ldots \rR_{n_2}[A]\,\rR_{n_1}[A]\ A\ ,
	\end{equation}
	\begin{equation}
	{\bf n} \= (n_1,n_2,\ldots,n_s) \quad\with n_i\in\N \und \sum_i n_i \= n\ ,
	\end{equation}
	where $1\le s \le n$, the $n{=}0$ term is the identity and
	\begin{equation}
	c_{\bf n} \= (-1)^s
	\bigl[ n_1\cdot(n_1+n_2)\cdots(n_1+n_2+\ldots+n_s)\bigr]^{-1} \ .
	\end{equation}
	Writing out the first few terms gives
	\begin{equation}
	\begin{aligned}
	T_gA &\= A \ -\ g\,\rR_1 A \ -\ \sfrac12g^2\bigl(\rR_2-\rR_1^2\bigr)A\ -\ 
	\sfrac16g^3\bigl(2\rR_3-\rR_1\rR_2-2\rR_2\rR_1+\rR_1^3\bigr)A \\[4pt]
	&\quad -\sfrac{1}{24}g^4\bigl(6\rR_4-2\rR_1\rR_3-3\rR_2\rR_2+\rR_1^2\rR_2-6\rR_3\rR_1
	+2\rR_1\rR_2\rR_1+3\rR_2\rR_1^2-\rR_1^4\bigr)A \ +\ {\cal O}(g^5)\ .
	\end{aligned}
	\end{equation}

	\noindent\textbf{Conventions and notation.} We work with the mostly plus metric $\eta^{\mu\nu}\=\mathrm{diag}(-1,+1,...,+1)$ and the Clifford algebra $\{\gamma^\mu, \gamma^\nu\}\=-2\eta^{\mu\nu}$.
	We generally suppress color indices and position labels, adopting the shorthand notations from section 4 in \cite{ALMNPP}. All objects are multiplied as color matrices or vectors, and integration kernels are convoluted with insertions of $A$.
	In the following we summarize the quantities that appear in the coupling flow operator. The fermionic propagator $S$ is the Green's function of the covariant derivative $\rD_\mu\=\partial_\mu+gA_\mu\times\ $ contracted with the gamma matrices (we also suppress Majorana spinor indices $\alpha, \beta,\ldots$),
	\begin{equation}
		S\ =\ \slashed{\rD}^{-1} \= - \bcontraction{}{\lambda}{\blambda}{\ }\lambda\ \blambda\ ,
	\end{equation}
	whereas the ghost propagator is given by
	\begin{equation}
		G \= \bigl(\sfrac{\partial \G(A)}{\partial A_\mu}\rD_\mu\bigr)^{-1}\=-\I\;\bcontraction{}{C}{\bC}{\; }C\ \bC\ .
	\end{equation}
	They can be expanded in the coupling as
	\begin{equation}
		\begin{aligned}
			&S\=S_0-gS_0\slA S\= \sum_{l=0}^{\infty}(-gS_0 \slA)^l S_0\=S_0 \sum_{l=0}^{\infty}(-g \slA S_0)^l\ ,\\
			&G\=G_0-gG_0 \sfrac{\partial \G(A)}{\partial A_\mu} A_\mu G\=\sum_{k=0}^{\infty}\Bigl(-gG_0 \sfrac{\partial \G(A)}{\partial A_\mu}A_\mu\Bigr)^k G_0\=G_0 \sum_{k=0}^{\infty}\Bigl(-g\sfrac{\partial \G(A)}{\partial A_\mu}A_\mu G_0\Bigr)^k\ ,
		\end{aligned}
	\end{equation}
	in terms of their free ($g=0$) versions 
	\begin{equation}
		G_0\=\bigl(\sfrac{\partial \G(A)}{\partial A_\mu}\partial_\mu\bigr)^{\;-1}\ ,\qquad
		S_0\=\slashed{\partial}^{\;-1}\=-\slashed{\partial} C\ ,\qquad
		C\=\Boxi\ ,
	\end{equation}
	with the scalar\footnote{It should be noted that we differ in our convention of $C$ by a minus sign compared to other texts, since we prefer $\Box C=1$.} propagator $C$.
	
	\newpage
	
	\noindent\textbf{Projectors and gauge-field decomposition.} It is useful to introduce the covariant gauge projector
	\begin{equation}\label{eq:cov_projector}
	P\indices{_\mu^\nu}\=\delta\indices{_\mu^\nu}-\rD_\mu G \sfrac{\partial \G(A)}{\partial A_\nu}\=\Pi\indices{_\mu^\sigma}\Bigl\{\delta\indices{_\sigma^\nu}-gA_\sigma \sum_{k=0}^{\infty}\Bigl(-gG_0 \sfrac{\partial \G(A)}{\partial A_\mu}A_\mu\Bigr)^k G_0 \sfrac{\partial \G(A)}{\partial A_\nu}\Bigr\}\ ,
	\end{equation}
	which in the free limit reduces to the projector
	\begin{equation}
		\Pi\indices{_\mu^\nu}\=\delta\indices{_\mu^\nu}-\partial_\mu G_0 \sfrac{\partial \G(A)}{\partial A_\nu}\ .
	\end{equation}
	They obey the identities
	\begin{equation}
		\sfrac{\partial \G(A)}{\partial A_\mu}\;P\indices{_\mu^\nu}\=0\=
		\sfrac{\partial \G(A)}{\partial A_\mu}\;\Pi\indices{_\mu^\nu}\ \qquad\textrm{and}\qquad
		P\indices{_\mu^\nu}\  \rD_\nu\=0\=
		\Pi\indices{_\mu^\nu}\; \partial_\nu\ .
	\end{equation}

	A decisive advantage of the Lorenz gauge $\sfrac{\partial \G(A)}{\partial A_\nu}\=\partial_\nu$ is that $G_0\ \equiv\ C$ so that $\Pi$ is equivalent to the standard transversal projector
	\begin{equation}
		\amalg\indices{_\mu^\nu}\=\delta\indices{_\mu^\nu}-\partial_\mu C \partial^\nu\ ,
	\end{equation}
	which fulfills the relations
	\begin{equation}
		\amalg\indices{_\mu^\rho}\ \Pi\indices{_\rho^\nu}\=\amalg\indices{_\mu^\nu}\ ,\qquad
		\Pi\indices{_\mu^\rho}\ \amalg\indices{_\rho^\nu}\=\Pi\indices{_\mu^\nu}\ ,
	\end{equation}
	and splits the Yang--Mills fields into transversal and longitudinal components
	\begin{equation}
		A_\mu\=\AT_\mu+\AL_\mu\ ,\qquad \AT_\mu\=\amalg\indices{_\mu^\nu}A_\nu\ ,\qquad \AL_\mu\=(\delta\indices{_\mu^\nu}-\amalg\indices{_\mu^\nu})A_\nu\=\partial_\mu C\ \partial\cdot A\ .
	\end{equation}
	For further convenience, we abbreviate $\partialA=\partial^\mu A_\mu$ and $B=C\;\;\partialA$.
	The longitudinal component of the gauge field $\AL_\mu$ lies in the kernel of both projectors,
	\begin{equation}\label{eq:kernel}
		\amalg\indices{_\mu^\nu}\AL_\nu\=\Pi\indices{_\mu^\nu}\AL_\nu\=0\ .
	\end{equation}
	Note that on the Landau gauge hypersurface ($\partialA=0$, $\xi\rightarrow 0$) the longitudinal component of the gauge field vanishes while $\AT\ \equiv\ A$. In general gauges (or outside of the gauge hypersurface of the Landau gauge), this is not the case. There, it turns out to be useful to define the `conjugate' Yang--Mills field
	\begin{equation}\label{eq:conjugate_fields}
		A^*_\mu\ :=\ \AT_\mu-\AL_\mu\= A_\mu -2\partial_\mu B\ ,
	\end{equation}
	related to $A_\mu$ by an involution
	\begin{equation}
		\Omega\indices{_\mu^\nu}\ :=\ \amalg\indices{_\mu^\nu}-(\delta\indices{_\mu^\nu}-\amalg\indices{_\mu^\nu})\ ,\qquad A^*_\mu\=\Omega\indices{_\mu^\nu}A_\nu\ .
	\end{equation}
	It satisfies various helpful properties like $\partial^\mu A_\mu^*\=-\partial^\mu A_\mu$ and $\partial_{[\mu}A_{\nu]}\=\partial_{[\mu}A^*_{\nu]}$. For the explicit construction of the Nicolai map, it is convenient to further define
	\begin{equation}
		{\Pi^*_\mu}^\nu\=\Omega\indices{_\mu^\rho}\ \Pi\indices{_\rho^\lambda}\ \Omega\indices{_\lambda^\nu}\=\Omega\indices{_\mu^\rho}\ \Pi\indices{_\rho^\nu}\=2\amalg\indices{_\mu^\nu}-\Pi\indices{_\mu^\nu}\ ,
	\end{equation}
	which is also a projector. Note the important simplification $\Pi^*\equiv\ \Pi$ for the Lorenz gauge.
	
	\section{Coupling flow operator}\label{sec:coupling_flow}
	\textbf{Rescaled construction.} In the canonical construction of the coupling flow operator, the $g$ derivative of the action is written as a supervariation. This requires that the supersymmetry is realized off-shell, in which case the action is the highest component of a superfield. In the following we therefore restrict ourselves to $D=4$. However, at least in the Landau gauge, the Nicolai map generalizes \cite{ALMNPP} to the critical dimensions $D=3,4,6,10$ \cite{BSS}. In gauge theories, the $g$ derivative of $S_{\textsc{SUSY}}$ cannot be written as a supervariation. A convenient solution is a rescaling of the field content with a suitable power of $g$ \cite{L1},
	such that the dependence on the coupling presents itself only as an overall factor in front of the action (except for a factor of $g$ multiplying the ghost term):
	\begin{equation}
		\begin{aligned}
			&S_{\textsc{SUSY}}[\tA, \tlambda, \tD, \tC, \tbC]\=S_{\mathrm{inv}}[\tA, \tlambda, \tD]\ +\ S_{\mathrm{gf}}[\tA, \tC, \tbC]\ ,\\[4pt]
			&S_{\mathrm{inv}}[\tA, \tlambda, \tD]\=\sfrac{1}{g^2}\intdx\;  \Bigl\{-\sfrac{1}{4}\tF^{\mu\nu}\tF_{\mu\nu}-\sfrac{\I}{2}\tblambda \slashed{\trD}\tlambda\;+\sfrac{1}{2}\tD^2\Bigr\}\ ,\\
			&S_{\mathrm{gf}}[\tA, \tC, \tbC]\=\sfrac{1}{g^2}\intdx\;\Bigl\{-\sfrac{1}{2\xi} \G(\tA)^2+g\,\tbC \sfrac{\partial \G(\tA)}{\partial \tA_\mu}\trD_\mu \tC\Bigr\}\ ,\\
		\end{aligned}
	\end{equation}
	where quantities with a tilde are rescaled. In particular, $\tA\=gA$ such that
	\begin{equation}
		\trD_\mu\=\partial_\mu+\tA_\mu \times \qquad \textrm{and}\qquad \tF_{\mu\nu}\=\partial_{\mu} \tA_{\nu}-\partial_{\nu} \tA_{\mu}+\tA_\mu\times \tA_\nu\ .
	\end{equation}
	In this scaling we can write the $g$ derivative of the action as a supervariation up to a Slavnov variation,
	\begin{equation}
		\partial_g S_{\textsc{SUSY}}\=-\sfrac{1}{g^3}\bigr\{\delta_\alpha \Delta_\alpha[\tA, \tlambda, \tD]-\sqrt{g}\,s\,\Delta_{\mathrm{gh}}[\tbC, \tA]\bigr\}\ ,
	\end{equation}
	with the superfield component
	\begin{equation}\label{eq:Delta_alpha}
		\Delta_\alpha[\tA, \tlambda, \tD]\=\sfrac{1}{4}\intdx\;  \Bigl\{-\sfrac{1}{2}\gamma^{\mu\nu}\;\; \tlambda\;\; \tF_{\mu\nu}\ +\ \gamma_5\;\; \tlambda\;\; \tD\Bigr\}_\alpha\ ,
	\end{equation}
	the ghost contribution
	\begin{equation}\label{eq:Delta_gh}
	\Delta_{\mathrm{gh}}[\tbC, \tA]\=\intdx\;  \Bigl\{\tbC\ \G(\tA)\Bigr\}\ ,
	\end{equation}
	the supervariations
	\begin{equation}\label{eq:supervariations}
	\delta_\alpha \tA_\nu\=-\I(\tblambda\gamma_\nu)_\alpha\ ,\qquad
	\delta_\alpha \tlambda_\beta\=-\sfrac{1}{2}(\gamma^{\mu\nu})_{\beta\alpha}\tF_{\mu\nu}-\tD(\gamma_5)_{\beta\alpha}\ ,\qquad
	\delta_\alpha \tD\=\I(\trD_\mu \tblambda \gamma_5 \gamma^\mu)_{\alpha} \ ,
	\end{equation}
	and the Slavnov variations
	\begin{equation}\label{eq:slavnov_variations}
	\begin{aligned}
	&s\tA_\mu\=\sqrt{g}\;\trD_\mu \tC\ ,\qquad&&s\tlambda\=\sqrt{g}\;\tlambda\times \tC \ ,\qquad&&s\tblambda\=\sqrt{g} \;\tblambda\times \tC \ ,\qquad&\\[4pt]
	&s\tD\=\sqrt{g}\;\tD\times \tC\ ,&&s\tC\=-\sfrac{\sqrt{g}}{2} \;\tC\times \tC\ ,&&s\tbC\=\sfrac{1}{\sqrt{g}}\sfrac{1}{\xi}\;\G(\tA)\ .&
	\end{aligned}
	\end{equation}
	It is convenient to integrate out the auxiliaries (with equation of motion $D=0$), so that
	\begin{equation}\label{eq:Delta_def}
	\Delta_\alpha[\tA, \tlambda]\=-\sfrac{1}{8} \intdx\; \Bigl\{\gamma^{\mu\nu}\;\; \tlambda\;\; \tF_{\mu\nu}\Bigr\}_\alpha\;.
	\end{equation}
	An intermediate rescaled coupling flow operator is then given by~\cite{L1,DL1}
	\begin{equation} \label{gaugeR}
	\R[\A] \= -\im\,\bcontraction{}{\Delta}{_\alpha[\A]\ }{\delta} \Delta_\alpha[\A]\ \delta_\alpha 
	+\sfrac{\im}{\sqrt{g}}\,\bcontraction{}{\Delta}{_{\textrm{gh}}[\A]\ }{s} \Delta_{\textrm{gh}}[\A]\ s
	-\sfrac{1}{\sqrt{g}}\,\bcontraction{}{\Delta}{_\alpha[\A]\ \bigl(}{\delta}  \Delta_\alpha[\A]\ \bigl(\delta_\alpha 
	\bcontraction{}{\Delta}{_{\textrm{gh}}[\A]\bigr)\ }{s} \Delta_{\textrm{gh}}[\A]\bigr)\ s\ .
	\end{equation}
	One can perform the contractions explicitly, obtaining gaugino and ghost propagators. Acting to the left, we write in compact notation
	\begin{equation}\label{eq:coupling_flow_operator}
	\begin{aligned}
	\tRl[\tA]&\= 
	-\sfrac{1}{8}\; \stackrel{\leftarrow}{\sfrac{\delta}{\delta \tA_\mu}}\tP\indices{_\mu^\nu}[\tA]\ \Tr\bigl(\gamma_\nu \tS[\tA]\,  \gamma^{\rho\lambda}\bigr)\tF_{\rho\lambda}+\stackrel{\leftarrow}{\sfrac{\delta}{\delta \tA_\mu}}\trD_\mu \tG[\A]\;\G(\A)\\
	&\=\stackrel{\leftarrow}{\sfrac{\delta}{\delta \tA_\mu}}\tA_\mu-\sfrac{1}{8}\; \stackrel{\leftarrow}{\sfrac{\delta}{\delta \tA_\mu}}\tP\indices{_\mu^\nu}[\tA]\ \Tr\bigl(\gamma_\nu \tS[\tA]\,  \bigl[2\partial\cdot \tA-\sltA \times\sltA\bigr]\bigr)\\
	&\=\stackrel{\leftarrow}{\sfrac{\delta}{\delta A_\mu}}A_\mu-\sfrac{1}{8}\; \stackrel{\leftarrow}{\sfrac{\delta}{\delta A_\mu}}P\indices{_\mu^\nu}[A]\ \Tr\bigl(\gamma_\nu S[A]\,  \bigl[2\partial\cdot A-g\slA \times\slA\bigr]\bigr)\ ,
	\end{aligned}
	\end{equation}
	where the first two lines are related by our choice of a linear 
	gauge-fixing function $\G(\A)$ and 
	by the identity
	\begin{equation}
	\gamma^{\rho\lambda}\tF_{\rho\lambda}\=2\slashed{\trD} \sltA +2\partial\cdot \tA-\sltA\times \sltA\ ,
	\end{equation}
	and we have inserted $\tA\=gA$ in the third line. 
	
	\noindent \textbf{Unrescaled coupling flow.} In \cite{LR} we demonstrated that the original (unrescaled) coupling flow operator is given by
	\begin{equation}
		R_g[A] \= \sfrac1g\,\bigl( \R[\tA] -E \bigr)\qquad \textrm{with}\qquad E \= A\,\sfrac{\delta}{\delta A}\ .
	\end{equation}
	Hence, we can write with $2\;S_0\;\partial\cdot A\=-2\slAL\=\slA^*-\slA$:
	\begin{equation}
		\begin{aligned}
			g\stackrel{\leftarrow}{R_g}[A]\=
			-&\sfrac{1}{8}\; \stackrel{\leftarrow}{\sfrac{\delta}{\delta A_\mu}}P\indices{_\mu^\nu}\ \Tr\;\bigl(\gamma_\nu S\,  \bigl[2\partial\cdot A-g\slA \times\slA\bigr]\bigr)\\[4pt]
			=-&\sfrac{1}{8}\; \stackrel{\leftarrow}{\sfrac{\delta}{\delta A_\mu}}P\indices{_\mu^\nu}\ \Tr\;\Bigl\{\gamma_\nu \Bigl[\sum_{l=0}^{\infty}(-gS_0\slA)^l\times(\slA^*-\slA)-\sum_{l=0}^{\infty}(-gS_0\slA)^lS_0 g\slA\times \slA\Bigr]\Bigr\}\\
			\=-&\sfrac{1}{8}\; \stackrel{\leftarrow}{\sfrac{\delta}{\delta A_\mu}}P\indices{_\mu^\nu}\ \Tr\;\Bigl\{\gamma_\nu \Bigl[\sum_{l=0}^{\infty}(-gS_0\slA)^l\times(\slA^*-\slA)+\sum_{l=1}^{\infty}(-gS_0\slA)^l \times\slA\Bigr]\Bigr\}\\
			\=
			-&\sfrac{1}{8}\; \stackrel{\leftarrow}{\sfrac{\delta}{\delta A_\mu}}P\indices{_\mu^\nu}\ \Tr\;\Bigl\{\gamma_\nu \Bigl[\sum_{l=1}^{\infty}(-gS_0\slA)^l\times(\slA^*-\slA)+\sum_{l=1}^{\infty}(-gS_0\slA)^l \times\slA\Bigr]\Bigr\}
			+\sfrac{1}{4}\; \stackrel{\leftarrow}{\sfrac{\delta}{\delta A_\mu}}P\indices{_\mu^\nu}\ \Tr\Bigl\{\gamma_\nu\slAL\Bigr\}\\
			\=-&\sfrac{1}{8}\; \stackrel{\leftarrow}{\sfrac{\delta}{\delta A_\mu}}P\indices{_\mu^\nu}\ \Tr\;\Bigl\{\gamma_\nu \sum_{l=1}^{\infty}(-gS_0\slA)^l\times\slA^*\Bigr\}
			- \stackrel{\leftarrow}{\sfrac{\delta}{\delta A_\mu}}P\indices{_\mu^\nu}\AL_\nu\\
			\=
			+&\sfrac{g}{8}\; \stackrel{\leftarrow}{\sfrac{\delta}{\delta A_\mu}}P\indices{_\mu^\nu}\ \Tr\;\Bigl\{\gamma_\nu \sum_{l=0}^{\infty}(-gS_0\slA)^lS_0\slA\times\slA^*\Bigr\}
			- \stackrel{\leftarrow}{\sfrac{\delta}{\delta A_\mu}}P\indices{_\mu^\nu}\AL_\nu\\
			\=
			+&\sfrac{g}{8}\; \stackrel{\leftarrow}{\sfrac{\delta}{\delta A_\mu}}P\indices{_\mu^\nu}\ \Tr\;\Bigl\{\gamma_\nu S\; \slA\times\slA^*\Bigr\}
			- \stackrel{\leftarrow}{\sfrac{\delta}{\delta A_\mu}}P\indices{_\mu^\nu}\AL_\nu\ .\\
		\end{aligned}
	\end{equation}
	The necessity that $R_g[A]$ contains no term of order $\mathcal{O}(g^{-1})$ directly follows from $\Pi\indices{_\mu^\nu}\AL_\nu\=0$ \eqref{eq:kernel}. More explicitly, we can expand the covariant projector \eqref{eq:cov_projector} to find
	\begin{equation}
		\begin{aligned}
			P\indices{_\mu^\nu}\AL_\nu
			\=&\Pi\indices{_\mu^\sigma}\Bigl\{\delta\indices{_\sigma^\nu}-gA_\sigma \sum_{k=0}^{\infty}\Bigl(-gG_0 \sfrac{\partial \G(A)}{\partial A_\rho}A_\rho\Bigr)^k G_0 \sfrac{\partial \G(A)}{\partial A_\nu}\Bigr\}\AL_\nu\\
			\=&-g\Pi\indices{_\mu^\sigma} A_\sigma G\ \sfrac{\partial \G(A)}{\partial A_\nu}\AL_\nu
			\=-g\Pi\indices{_\mu^\sigma} A_\sigma \sum_{k=0}^{\infty}\Bigl(-gG_0 \sfrac{\partial \G(A)}{\partial A_\rho}A_\rho\Bigr)^k \times B\ .
		\end{aligned}
	\end{equation}
	Therefore, the coupling flow generator can be written succinctly as
	\begin{equation}\label{eq:R_compact}
		\stackrel{\leftarrow}{R_g}[A]\=
		\sfrac{1}{8}\; \stackrel{\leftarrow}{\sfrac{\delta}{\delta A_\mu}}P\indices{_\mu^\nu}\ \Tr\Bigl\{\gamma_\nu S\; \slA\times\slA^*\Bigr\}
		\ +\ \stackrel{\leftarrow}{\sfrac{\delta}{\delta A_\mu}}\Pi\indices{_\mu^\sigma} A_\sigma \;G\; \sfrac{\partial \G(A)}{\partial A_\nu}\AL_\nu\ ,
	\end{equation}
	or in its full perturbative expansion as
	\begin{equation}\label{eq:pert_expansion}
		\begin{aligned}
			\stackrel{\leftarrow}{R_g}[A]\=
			\sfrac{1}{8}\; &\stackrel{\leftarrow}{\sfrac{\delta}{\delta A_\mu}}\Pi\indices{_\mu^\sigma}\Bigl\{\delta\indices{_\sigma^\nu}-gA_\sigma G_0 \sum_{k=0}^{\infty}\Bigl(-g\sfrac{\partial \G(A)}{\partial A_\rho}A_\rho G_0\Bigr)^k  \sfrac{\partial \G(A)}{\partial A_\nu}\Bigr\}\ \Tr\Bigl\{\gamma_\nu S_0 \sum_{l=0}^{\infty}(-g \slA S_0)^l  \slA\times\slA^*\Bigr\}\\
			+\ &\stackrel{\leftarrow}{\sfrac{\delta}{\delta A_\mu}}\Pi\indices{_\mu^\sigma} A_\sigma \sum_{k=0}^{\infty}\Bigl(-gG_0 \sfrac{\partial \G(A)}{\partial A_\rho}A_\rho\Bigr)^k \times B\ ,
		\end{aligned}
	\end{equation}
	which is the main result of this work. Together with \eqref{eq:nmap_coefficients}, equation~\eqref{eq:pert_expansion} offers a straightforward way to compute the Nicolai map in any gauge and to any order by collecting the powers of the gauge coupling. The first term generalizes\footnote{Moreover, a similar graphical representation with Feynman-like graphs is possible.} the Landau-gauge on-shell flow operator of \cite{ALMNPP}, with the distinct modification of the last factor $A$ to $A^*$. For our linear gauges, the second term reduces on the gauge hypersurface to
	\begin{equation}
		\stackrel{\leftarrow}{\sfrac{\delta}{\delta A_\mu}}A_\mu \times B\ ,
	\end{equation}
	but outside the gauge hypersurface it appears in even the Landau gauge. For gauge-invariant functionals $X[A]$, our coupling flow operator simplifies considerably. Such situations are left for further study.
	
	\section{Axial gauge}
	\textbf{Map construction.} As an example, we calculate the $D=4$ Nicolai map in the axial gauge up to the second order. This includes the light-cone gauge. Inserting $\G(A)\ =\ n\cdot A$, we obtain
	the coupling flow operator
	\begin{equation}\label{eq:pert_expansion_axial}
	\begin{aligned}
	\stackrel{\leftarrow}{R_g}[A]\=
	\sfrac{1}{8}\; &\stackrel{\leftarrow}{\sfrac{\delta}{\delta A_\mu}}\Pi\indices{_\mu^\sigma}\Bigl\{\delta\indices{_\sigma^\nu}-gA_\sigma G_0 \sum_{k=0}^{\infty}\Bigl(-g n\cdot A G_0\Bigr)^k  n^\nu \Bigr\}\ \Tr\Bigl\{\gamma_\nu S_0 \sum_{l=0}^{\infty}(-g \slA S_0)^l  \slA\times\slA^*\Bigr\}\\
	+&\stackrel{\leftarrow}{\sfrac{\delta}{\delta A_\mu}}\Pi\indices{_\mu^\sigma} A_\sigma \sum_{k=0}^{\infty}\Bigl(-gG_0 n\cdot A\Bigr)^k \times B\ ,
	\end{aligned}
	\end{equation}
	With $S_0\=-\slashed{\partial}C$, we can read off (see \eqref{eq:series_R})
	\begin{equation}\label{eq:R1_R2}
	\begin{aligned}
	\rRl_1\=&\stackrel{\leftarrow}{\sfrac{\delta}{\delta A_\mu}} \Pi\indices{_\mu^\sigma}\left[A_\sigma \times B-\frac{1}{8}\Tr\left(\gamma_\sigma\, \slashed{\partial} C\, \slA\times\slA^*\right)\right]\ ,\\
	\rRl_2\=&\stackrel{\leftarrow}{\sfrac{\delta}{\delta A_\mu}}\Pi\indices{_\mu^\sigma}\left[-A_\sigma G_0 n{\cdot} A {\times} B+\sfrac{1}{8}A_\sigma G_0 n_\nu \Tr\left(\gamma^\nu\, \slashed{\partial}C\, \slA {\times} \slA^*\right)-\sfrac{1}{8}\Tr\left(\gamma_\sigma\, \slashed{\partial}C\, \slA\  \slashed{\partial}C\, \slA {\times} \slA^*\right)\right]\ .
	\end{aligned}
	\end{equation}
	Evaluating the traces gives
	\begin{equation}\label{eq:trR1}
	\rR_1 A_\mu\=-\Pi\indices{_\mu^\nu}\Aone_\nu\ \ \with \Aone_\nu\ :=\ C^\rho A_{[\rho}A^{*}_{\nu]}+\sfrac{1}{2}C_\nu A^\rho A^*_\rho-A_\nu B\ , 
	\end{equation}
	and
	\begin{equation}\label{eq:R2}
		\begin{aligned}
			\rR_2 A_\mu\=\Pi\indices{_\mu^\nu}\Bigl[A_\nu G_0 n\cdot \Aone-3 C^\rho A^\lambda C_{[\nu}A_\rho A^*_{\lambda]}+2C^\rho A_{[\rho}\Aone_{\nu]}+2C^\rho A_{[\rho}A_{\nu]} B\Bigr]\ ,
		\end{aligned}
	\end{equation}
	where we further simplified notation by writing $\partial^\rho C\ \equiv\ C^\rho$ and with the understanding that the last object in each line is a color vector instead of a color matrix.
	For the second order we also need
	\begin{equation}
		\rR_1^2A_\mu\=\Pi\indices{_\mu^\nu}\Bigl[B\Pi\indices{_\nu^\lambda}\Aone_\lambda-A_\nu C^\sigma \Pi\indices{_\sigma^\lambda}\Aone_\lambda+C^\rho A_{[\rho}\Pi^{*\;\, \lambda}_{\nu]}\Aone_\lambda+C^\rho A^*_{[\rho}\Pi^{\hphantom{*}\;\, \lambda}_{\nu]}\Aone_\lambda\Bigr]\ ,
	\end{equation}
	which we can simplify by inserting $\Pi^*\=2\amalg-\Pi$ and identifying $(A-A^*)_\rho\=2\AL_\rho\=2\partial_\rho B$:
	\begin{equation}\label{eq:R1sq}
		\begin{aligned}
		\rR_1^2A_\mu
		\=&\Pi\indices{_\mu^\nu}\Bigl[
		B\Pi\indices{_\nu^\lambda}\Aone_\lambda
		-A_\nu C^\sigma \Pi\indices{_\sigma^\lambda}\Aone_\lambda
		+2C^\rho A_{[\rho}\amalg\indices{_{\nu]}^\lambda}\Aone_\lambda
		-2C^\rho \partial_{[\rho}B\ \Pi\indices{_{\nu]}^\lambda}\Aone_\lambda\Bigr]\\
		\=&\Pi\indices{_\mu^\nu}\Bigl[
		-A_\nu C^\sigma \Pi\indices{_\sigma^\lambda}\Aone_\lambda
		+2C^\rho A_{[\rho}\amalg\indices{_{\nu]}^\lambda}\Aone_\lambda
		+2C^\rho B\partial_{[\rho}\Aone_{\nu]}\Bigr]\ .
		\end{aligned}
	\end{equation}
	In the second step we have used integration by parts\footnote{when integrating by parts away from $B$ we obtain two terms: One where the derivative moves to the left, without a minus sign due to the structure of the position labels and one where the derivative moves to the right with a sign flip.} in the last term and used
	\begin{equation}
		\Pi\indices{_\mu^\nu}C_\nu\=0\ ,\qquad C\indices{^\rho_\rho}\=1\ ,\qquad \partial_{[\rho}\Pi\indices{_{\nu]}^\lambda}\=\partial_{[\rho}\delta\indices{_{\nu]}^\lambda}\ .
	\end{equation} 
	With 
	\begin{equation}
		C^\sigma\Pi\indices{_\sigma^\lambda}\=C^\lambda-G_0 n^\lambda\ ,
	\end{equation}
	when combining \eqref{eq:R2} and \eqref{eq:R1sq} to
	\begin{equation}
		\begin{aligned}
			\bigl(\rR_1^2A-\rR_2A\bigr)_\mu\=\Pi\indices{_\mu^\nu}\Bigl[
			-A_\nu C^\lambda \Aone_\lambda
			-2C^\rho A_{[\rho}C\indices{_{\nu]}^\lambda}\Aone_\lambda
			+2C^\rho B \partial_{[\rho}\Aone_{\nu]}
			+3C^\rho A^\lambda C_{\nu}A_\rho A^*_\lambda
			-2C^\rho A_{[\rho}A_{\nu]}B
			\Bigr]\ ,
		\end{aligned}
	\end{equation}
	we have found a structure where the ghost contribution only resides in the projector $\Pi$ at the beginning of each term.\footnote{It is left for further study whether this can be achieved at higher orders.} 
	Inserting $\Aone_\lambda$ of \eqref{eq:trR1}, 
	with relations
	\begin{equation}
		C_\lambda C^\lambda\= CC\indices{_\lambda^\lambda}\=C\ ,\qquad C^{[\rho\lambda]}\=0\ ,\qquad \textrm{etc.}
	\end{equation}
	one quickly finds the following expression for the Nicolai map $T_gA=A-g \rR_1 A+\frac{g^2}{2}(\rR_1^2-\rR_2)A+\mathcal{O}(g^3)$ in the axial gauge for $D=4$:
	\begin{equation}\label{eq:nmap_axial}
		\begin{aligned}
			T_gA_\mu\=A_\mu+g\Pi\indices{_\mu^\nu}\bigl\{C^\rho &A_{[\rho}A^{*}_{\nu]}-A_\nu B\bigr\}\\
			+\sfrac{g^2}{2} \Pi\indices{_\mu^\nu}\bigl\{
			3&C^\rho A^\lambda C_{[\nu}A_\rho A^*_{\lambda]}
			-2C^\rho A_{[\rho}A_{\nu]}B
			-2C^\rho B\partial_{[\rho}(A_{\nu]}B)\\
			+2&C^\rho A_{[\rho}C\indices{_{\nu]}^\lambda}A_\lambda B
			-C^\rho  A_{[\rho}C_{\nu]}A^\lambda A^*_\lambda
			+A_\nu C^\lambda A_\lambda B
			-\sfrac{1}{2}A_\nu C A^\lambda A^*_\lambda
			\bigr\}\\
			+g^2\Pi\indices{_\mu^{[\nu}}C^{\rho]}&B C\indices{_\rho^\sigma}A_{[\sigma}A^*_{\nu]}\ +\ \mathcal{O}(g^3)\ .
		\end{aligned}
	\end{equation}
	Note that when setting $B\=0$, $A\=A^*$ and $\Pi\=\amalg$, one recovers the known expression for the Landau gauge (on its gauge hypersurface). From $n^\mu \Pi_\mu^{\ \nu}=0$ it follows trivially that this map 
	respects the gauge condition $n\cdot T_g A\=0$. The free-action and 
	determinant-matching conditions,
	\begin{equation}
		S_0[A']\=S_g[A] \qquad\textrm{and}\qquad \det\Bigl(\sfrac{\delta A'}{\delta A}\Bigr)\= \MSS\ \FP\ ,
	\end{equation}
	must also be obeyed, as they follow from the general construction. They 
	may be verified as a cross-check of the expression in \eqref{eq:nmap_axial}.

	\section{Conclusions and outlook}
	For ${\cal N}{=}\,1$ off-shell supersymmetric Yang--Mills theory with 
	arbitrary gauge fixing,
	we have recast the coupling flow operator in a compact form with the 
	help of the gauge projector~$\Pi$
	and a `conjugate' gauge field $A^*=A^{\mathrm{T}}-A^{\mathrm{L}}$.
	This allows for an efficient construction of the Nicolai map even off 
	the gauge hypersurface,
	as we demonstrated for the axial gauge to second order in the coupling.
	It also elucidates the special character of the Landau gauge.
	Our formulation should also facilitate applications of the Nicolai map, 
	like the ones initiated in \cite{DL2, NP}.
	
	Various further investigations come to mind, such as a study of the 
	light-cone gauge,
	the effect of gauge transformations on the Nicolai map, a graphical 
	representation of its
	perturbative expansion, its convergence properties, possible 
	non-uniqueness related to
	the kernel of the coupling flow operator, a chiral variant of the map, 
	an on-shell extension
	to higher dimensions and, not the least, the case of extended 
	supersymmetry, in particular
	of ${\cal N}{=}\,4$ super Yang--Mills theory in four dimensions.
	
	When writing up the present paper we received the preprint \cite{MN}, which 
	has substantial overlap
	with ours, but organizes the coupling 
	flow operator in a different fashion. We have checked that their result for the axial gauge agrees  with ours.
	
	\bigskip
	\noindent
	{\bf Acknowledgment.\ } 
	M.R.~is supported by a PhD grant of the German Academic Scholarship Foundation.


\newpage 
	
	\begin{thebibliography}{99}
		\addtolength{\itemsep}{-3.5pt}
		\bibitem{Nic1}
		H.~Nicolai,
		{\it On a new characterization of scalar supersymmetric theories},\\
		\href{https://dx.doi.org/10.1016/0370-2693(80)90138-0}
		{{\it Phys.\ Lett.\ B} {\bf 89} (1980) 341}.
		\bibitem{Nic2}
		H.~Nicolai,
		{\it Supersymmetry and functional integration measures},\\
		\href{https://dx.doi.org/10.1016/0550-3213(80)90460-5}
		{{\it Nucl.\ Phys.\ B} {\bf176} (1980) 419}.
		\bibitem{Nic3}
		H.~Nicolai,
		{\it Supersymmetric functional integration measures},\\
		lectures delivered at the NATO Advanced Study Institute on Supersymmetry,\\
		Bonn, Germany, 20--31 Aug 1984, 
		\href{https://cds.cern.ch/record/155731?ln=en}
		{pp.393--420, eds. K.~Dietz et. al., {\it Plenum Press} (1984)}.
		\bibitem{LR}
		O.~Lechtenfeld and M.~Rupprecht,
		{\it Universal form of the Nicolai map},\\
		Preprint.
		[\href{https://arxiv.org/abs/2104.00012}{arXiv:2104.00012 [hep-th]}].
		\bibitem{ALMNPP}
		S.~Ananth, O.~Lechtenfeld, H.~Malcha, H.~Nicolai, C.~Pandey and S.~Pant,
		{\it Perturbative linearization of supersymmetric Yang--Mills theory},
		\href{https://dx.doi.org/10.1007/JHEP10(2020)199}
		{{\it JHEP} {\bf 10} (2020) 199}
		[\href{https://arxiv.org/abs/2005.12324}{arXiv:2005.12324 [hep-th]}].
		\bibitem{BSS}
		L. Brink, J.~H. Schwarz and J.~Scherk,
		{\it Supersymmetric Yang-Mills theories},\\
		\href{https://doi.org/10.1016/0550-3213(77)90328-5}
		{{\it Nucl.\ Phys.\ B} {\bf121} (1977) 77}.
		\bibitem{L1}
		O.~Lechtenfeld,
		{\it Construction of the Nicolai mapping in supersymmetric field theories},\\
		Ph.D.\ Thesis, Bonn University (1984),
		\href{https://lib-extopc.kek.jp/preprints/PDF/2000/0030/0030157.pdf}
		{internal report {\it BONN-IR-84-42}, ISSN-0172-8741}.
		\bibitem{DL1}
		K.~Dietz and O.~Lechtenfeld,
		{\it Nicolai maps and stochastic observables from a coupling constant flow},
		\href{https://dx.doi.org/10.1016/0550-3213(85)90132-4}
		{{\it Nucl.\ Phys.\ B} {\bf 255} (1985) 149}.
		\bibitem{DL2}
		K.~Dietz and O.~Lechtenfeld,
		{\it Ghost-free quantisation of non-Abelian gauge theories 
			via the Nicolai transformation of their supersymmetric extensions},
		\href{https://dx.doi.org/10.1016/0550-3213(85)90642-X}
		{{\it Nucl.\ Phys.\ B} {\bf 259} (1985) 397}.
		\bibitem{NP}
		H.~Nicolai and J.~Plefka,
		{\it ${\cal N}{=}\,4$ super-Yang--Mills correlators without anticommuting variables},\\
		\href{https://dx.doi.org/10.1103/PhysRevD.101.125013}
		{{\it Phys.\ Rev.\ D} {\bf 101} (2020) 125013}
		[\href{https://arxiv.org/abs/2003.14325}{arXiv:2003.14325 [hep-th]}].
		\bibitem{MN}
		H.~Malcha and H.~Nicolai,
		{\it Perturbative linearization of super-Yang-Mills
			theories in arbitrary gauges},\\
		Preprint.
		[\href{https://arxiv.org/abs/2104.06017}{arXiv:2104.06017 [hep-th]}].
	\end{thebibliography}
\end{document}